\documentclass[twocolumn, showpacs,pra,floatfix,eqsecnum]{revtex4}

\usepackage{graphicx}
\usepackage{bm}
\usepackage{amsmath}
\usepackage{amssymb}
\usepackage{upgreek}

\newcommand{\be}{\begin{equation}}
\newcommand{\ee}{\end{equation}}

\newcommand{\rmA}{\mathrm{A}}
\newcommand{\AM}{\mathrm{AM}}

\begin{document}

\title{On the Possibility of Measuring the Abraham Force using Whispering Gallery Modes}
\author{I. Brevik and S. {\AA}. Ellingsen}
\address{Department of Energy and Process Engineering, Norwegian University of Science and Technology, N-7491 Trondheim, Norway}
\email{iver.h.brevik@ntnu.no} %% email address is required

\begin{abstract}

Critical experimental tests of the time-dependent Abraham force in phenomenological electrodynamics are scarce.
 In this paper we analyze the possibility of making use of intensity-modulated whispering gallery modes in a microresonator
  for this purpose. Systems of this kind appear attractive, as the strong concentration of electromagnetic fields near
   the rim of the resonator serves to enhance the Abraham torque exerted by the field. We analyze mainly  spherical resonators,  although
     as an introductory step we consider also the cylinder geometry. The order of magnitude of the Abraham torques are estimated by inserting
     reasonable and   common values for the various input parameters. As expected, the predicted torques turn out to be very small,
      although probably not beyond any reach experimentally. Our main idea is essentially a generalization  of the method  used by G. B. Walker
      {\it et al.}\ [Can. J. Phys. {\bf 53}, 2577 (1975)] for low-frequency fields, to the optical case.
\end{abstract}


\maketitle

%%%%%%%%%%%%%%%%%%%%%%%%%%%%%%%%%
%%%%%%%% S E C T I O N %%%%%%%%%%
%%%%%%%%%%%%%%%%%%%%%%%%%%%%%%%%%
\section{Introduction}

The one-hundred years old Abraham-Minkowski energy-momentum problem in
phenomenological electrodynamics \cite{abraham09,minkowski10} has
recently attracted considerable interest. Assume  henceforth for
simplicity that the medium is nonmagnetic and nondispersive, with
refractive index $n$.  In our opinion -- as expressed in the review
article some years ago by one of the present authors
\cite{brevik79} -- the most physical expression for the
electromagnetic force density is the Abraham expression (SI units
assumed)
\begin{equation}
{\bf f}^\rmA={\bf f}^\AM +\frac{n^2-1}{c^2} \frac{\partial}{\partial
t}\left(\bf E \times H \right). \label{1A}
\end{equation}
Here the first term $ {\bf f}^\AM =-(\epsilon_0/2)E^2{\bf
\nabla}n^2$ is different from zero in regions where $n$ varies
with position, especially in the surface regions of dielectrics.
This term is common for the Abraham and Minkowski tensors, and may
appropriately be called the Abraham-Minkowski term. The second,
time-dependent term in Eq.~(\ref{1A}), is the Abraham term.  It
may be noted that the expression (\ref{1A}) is in agreement with
Ginzburg \cite{ginzburg79}, as well as with Landau and Lifshitz
\cite{landau84}.

One may ask: is it possible to detect the Abraham term in
experiment? The answer is yes, but the task has proven to be
surprisingly difficult. The magnitude of the electromagnetic
frequency is a significant factor in this context. Let us give a
brief account of three important experimental cases:

1) The first case is the quasi-stationary torque experiment of
Walker {\it et al.}\ \cite{walker75,walker75a}. Strong,
time-varying, orthogonal electric and magnetic fields were applied
across a dielectric shell of high permittivity, making it possible
to detect the {\it oscillations themselves.} In this way the
Abraham term was measured quantitatively.

2) When considering instead high-frequency fields such as in
optics, the Abraham term fluctuates out when averaged over a
period. One can  thus no longer detect this force directly. The
physical effect of this force is however to produce an accompanying
mechanical momentum propagating together with the  Abraham
momentum. The resulting total momentum is the Minkowski momentum,
corresponding to the divergence-free Minkowski energy-momentum tensor. This tensor
has the particular property of being space-like, corresponding to the possibility of getting negative field energy in certain inertial frames. An authoritative
experiment measuring the Minkowski momentum is that of Jones {\it
et al.}\ \cite{jones54,jones78}, measuring the radiation pressure on a
mirror immersed in a dielectric liquid. Both cases 1)
and 2) are discussed in some detail in Ref.~\cite{brevik79}.

3) The third example to be mentioned is the photon recoil experiment of Campbell {\it et al.}\ \cite{campbell05}, where the photon momentum in a medium (in this case a Bose-Einstein condensate) was found to be equal to the Minkowski value $\hbar \bf k$.

Most other experiments are measuring not the Abraham term but rather the surface force ${\bf f}^\AM $, although claims are sometimes made to the contrary. In our opinion this is the case also for the interesting new fiber optical experiment of She {\it et al.}\ \cite{she08}; cf.\ the remarks in Refs.~\cite{brevik09,brevik10}.

Our main purpose in the present paper is however not to interpret already existing experiments, but instead to propose the idea of using {\it whispering gallery modes} as a convenient experimental tool to detect the Abraham term in optics. To our knowledge this idea has not been considered before. Whispering gallery modes are commonly produced in microspheres; they have a large circulating power, about 100 W typically, and the field energy is concentrated along the rim of the sphere. That means, if such a sphere is suspended in the gravitational field and fed with an appropriate intensity  modulated field, the sphere becomes exposed to a vertical torque according to Eq.~\eqref{1A}. With the field energy essentially concentrated along the rim, the arm in the torque calculation is essentially the same as the radius, thus maximizing the torque. In effect, this is the idea of the experiment of Walker {\it et al.}\ \cite{walker75,walker75}, generalized to optical frequencies.  We have actually suggested this idea qualitatively before, in Refs.~\cite{brevik09,brevik10}.

The next two sections give quantitative estimates for performing such an experiment. The torque turns out to be small, as expected, but not beyond any possibility for experimental detection. Spherical geometry, as mentioned, is most typical for the whispering gallery setup. In the next section we consider however as an introductory step the somewhat more simple geometry of a cylindrical shell.

Before closing this section, let us give a few more references to the Abraham-Minkowski problem, in addition to the references given above. A nice introduction can be found in M{\o}ller's book \cite{moller72}. A review, up to 2007, is given by Pfeifer {\it et al.}\ \cite{pfeifer07}. Some more recent papers are Refs.~\cite{brevik09a,barnett10,barnett10a}.

%%%%%%%%%%%%%%%%%%%%%%%%%%%%%%%%%
%%%%%%%% S E C T I O N %%%%%%%%%%
%%%%%%%%%%%%%%%%%%%%%%%%%%%%%%%%%
\section{Cylindrical geometry}
Consider first as the simplest case a compact cylinder of length $L$ and radius $a$. On the inside, $r<a$, the permittivity is $\epsilon$ and the permeability $\mu$. On the outside, $r>a$,  a vacuum is assumed.
The dispersion relation for stationary modes is known to be \cite{stratton41}
%\begin{widetext}
\begin{align}
&\left[ \frac{\mu}{u}\frac{J'_{m}(u)}{J_{m}(u)}-\frac{1}{v}
\frac{{H_{m}^{(1)}}'(v)}{H_{m}^{(1)}(v)}\right]
\left[\frac{\epsilon\,\omega^2}{u}\frac{J_{m}'(u)}{J_{m}(u)}-
\frac{\omega^2}{v}\frac{{H_{m}^{(1)}}'(v)}{H_{m}^{(1)}(v)}\right]\notag \\
&=m^2 k^2\left(\frac{1}{v^2}-\frac{1}{u^2}\right)^2. \label{1}
\end{align}
%\end{widetext}
We are working with SI units and let $\epsilon$ and $\mu$ be dimensional, so that ${\bf D}=\epsilon \bf E$, ${\bf B}=\mu \bf H$. The transverse wave vectors on the
inside and the outside are
\begin{equation}
\lambda_1=n\omega/c, \quad \lambda_2=\omega/c, \label{2}
\end{equation}
respectively, while their nondimensional counterparts are
\begin{equation}
u=\lambda_1a, \quad v=\lambda_2a. \label{3}
\end{equation}

An important property of this equation is that when the axial wave
vector $k=0$ -- as is of interest here as we we consider azimuthal
modes only -- the right-hand side vanishes and the problem becomes
separable into TE and TM modes.

We write the mode expansions for the fields in the inner region \cite{stratton41}:
\begin{subequations}
\begin{align}
E_r=&-\frac{\mu\omega}{\lambda_1^2r}\sum_{m=-\infty}^\infty
mJ_m(\lambda_1r)\,b_mF_m, \label{4}  \\
E_\theta=&-\frac{i\mu\omega}{\lambda_1}\sum_{m=-\infty}^\infty
J_m'(\lambda_1r)\,b_m F_m, \label{5}\\
E_z=&\sum_{m=-\infty}^\infty J_m(\lambda_1r)\,a_m F_m, \label{6}
\end{align}
\end{subequations}
and
\begin{subequations}
\begin{align}
H_r=&\frac{\epsilon\omega}{\lambda_1^2r}\sum_{m=-\infty}^\infty
mJ_m(\lambda_1r)\,a_mF_m, \label{7}\\
H_\theta=&\frac{i\epsilon\omega}{\lambda_1}\sum_{m=-\infty}^\infty
J_m'(\lambda_1r)\,a_mF_m, \label{8}\\
H_z=&\sum_{m=-\infty}^\infty J_m(\lambda_1r)\,b_mF_m, \label{9}
\end{align}
\end{subequations}
where
\begin{equation}
F_m=e^{im\theta-i\omega t}. \label{10}
\end{equation}
The coefficients $a_m$ and $b_m$, corresponding to the TM and TE modes, give the weight of each mode.

In our considerations below we will for simplicity extract one
single TE mode of high order $m$, such that there is an
azimuthally moving momentum  concentrated in the vicinity of the
boundary $r=a$. (In reality, the incident power may be distributed
over a band of neighbouring $m$ modes, but this does not influence
the essence of our argument.) We first need to determine the
magnitude of the radial argument $\lambda_1r \approx u$. Let us
take
\begin{equation}
m=100, \quad n=1.5, \quad a=100~ {\mu\rm m}. \label{11}
\end{equation}
It is known that for a large value of the order $m$ the first
maximum of the function $J_m(x)$ occurs when $x$ is very close to
$m$. This maximum is the one of interest here. Thus the lowest
resonance frequency $\omega$ is determined by the equation
\begin{equation}
na\omega /c=m. \label{11a}
\end{equation}
With the numbers given above,
 \begin{equation}
  \omega=2\times
10^{16}~\rm \text{s}^{-1}. \label{12}
\end{equation}
In this manner we manage to make the beam  strongly concentrated
near the rim, as desired. One has in this case $E_z=0$, $H_r=0$,
while the
 nonvanishing field components of interest are
\begin{align}
E_r=&-\frac{\mu\omega}{\lambda_1^2\,r}
\,mJ_m(\lambda_1r)\,b_mF_m, \label{13}\\
H_z=&J_m(\lambda_1r)b_mF_m. \label{14}
\end{align}
The azimuthal component of the Poynting vector ${\bf S}(r)$ in the interior is
\begin{equation}
S_\theta(r)=-\frac{1}{2}\Re [E_rH_z^*]=\frac{\mu\omega m}{2\lambda_1^2r}J_m^2(\lambda_1r)|b_m|^2, \label{15}
\end{equation}
corresponding to the azimuthal power
\begin{equation}
P= L\int_0^a S_\theta \,dr=\frac{\mu\omega mL}{2\lambda_1^2}|b_m|^2\int_0^u \frac{dx}{x}J_m^2(x). \label{16}
\end{equation}
In our case the factor $1/x$ can be extracted outside the integral, so that
\begin{equation}
P=\frac{\mu\omega mL}{2\lambda_1^2u}|b_m|^2\int_0^udx J_m^2(x). \label{17}
\end{equation}

Assume now that the beam is intensity modulated with a  frequency $\omega_0$  ($\omega_0$ low compared with optical frequencies),
\begin{equation}
P=P_0\cos \omega_0 t, \quad S_\theta=S_0 \cos \omega_0 t. \label{18}
\end{equation}
Then the azimuthal Abraham force density $f_\phi^\rmA$ is
\begin{equation}
f_\phi^\rmA=\frac{n^2-1}{c^2}\frac{\partial S_\theta}{\partial t}=-\frac{n^2-1}{c^2}\omega_0S_0\sin \omega_0t, \label{19}
\end{equation}
giving rise to the following Abraham torque $N_z^\rmA$ around the vertical symmetry axis:
\begin{equation}
N_z^\rmA=2\pi L\int_0^a r^2f_\phi^\rmA dr \approx 2\pi La^2\int_0^af_\phi^\rmA dr. \label{20}
\end{equation}
Defining the quantity $K$ as
\begin{equation}K=-\frac{n^2-1}{c^2}2\pi a^2P_0, \label{21}
\end{equation}
we thus see that the torque can be written as
\begin{equation}
N_z^\rmA=K\omega_0\sin \omega_0 t. \label{22}
\end{equation}
As expected, the torque becomes very small. As order of magnitude
we get
\begin{equation}
K \sim \frac{2\pi a^2}{c^2}P_0 \sim (0.7\times 10^{-24}\mathrm{s}^2)\cdot P_0
\label{23}
\end{equation}
and the Abraham torque is estimated as
\begin{equation}
N_z^\rmA \sim  (0.7\times 10^{-24}\mathrm{s}^2)\cdot \omega_0P_0.
\label{24}
\end{equation}
Insert first the very low value of $\omega_0 \sim 1~\rm s^{-1}$, and take
$P_0\sim 100~$W. We  get $N_z\sim 0.7\times 10^{-22}~\rm N~m$, which is
 much less than the value $10^{-16}~\rm N~m$ obtained in the
classic Beth experiment \cite{beth36}, for example, in which the angular momentum of
light was measured. It is however possible to improve the situation by exploiting the fact that the build-up and ringdown times for this kind of resonators are known to be very small, in the order of tens to hundreds of Ns (see discussion below). It is thus realistic to insert a much higher value for $\omega_0$. Inserting tentatively $\omega_0= 1000~\rm s^{-1}$ we get $N_z\sim 0.7\times 10^{-19}~\rm N~m$, which is perhaps not so unrealistic after all.

It is physically instructive to look at the system in another way, by considering the angular deflection $\phi$ of the cylinder instead of the magnitude of the torque. Let the cylinder be hanging vertically in the gravitational field, suspended by a thin wire of known torsion constant $\kappa$. Denoting the eigenfrequency of the cylinder in the absence of any torque by $\Omega$, and denoting the damping coefficient by $\gamma$, we have as  equation of motion
\begin{equation}
\ddot{\phi}+\gamma \dot{\phi}+\Omega^2\phi=\frac{K}{I}\omega_0\sin \omega_0t.
\label{25}
\end{equation}
Here $I=\frac{1}{2}Ma^2$ is the moment of inertia about the $z$ axis, $M=\rho a^2 L$ being the cylinder mass with $\rho$ the material density. In our notation, $\kappa=I\Omega^2$. With $a=100~\upmu$m as above we obtain, when choosing $L=1~$mm and assuming  $\rho \sim 10^3~\rm kg/m^3$,
\begin{equation}
\Omega=\sqrt{\kappa/I} \sim 10^8\sqrt{\kappa}. \label{25A}
\end{equation}
 For the magnitude of $\kappa$  we may choose a typical value characteristic for  torsion experiments testing the equivalence principle, $\kappa \sim 10^{-9}~\rm N~m/rad $ \cite{hou03,schlamminger08}. Then,
 \begin{equation}
 \Omega \sim 10^3 \rm ~rad~s^{-1}. \label{25B}
 \end{equation}
 The magnitude of $\Omega$ is large because $a$ is assumed small.

 The largest oscillations  occur at resonance, when $\omega_0$ is chosen equal to $\Omega$. Then,
\begin{equation}
\phi=-\frac{K}{I\gamma}\cos \Omega t. \label{26}
\end{equation}
The maximum value, when $P_0 \sim 100~$W, is
\begin{equation}
\phi_\mathrm{max}=\frac{n^2-1}{c^2}\frac{4}{M}\frac{P_0}{\gamma} \sim \frac{10^{-7}}{\gamma} \rm \frac{rad}{s}. \label{27}
\end{equation}
It would be of interest to make an estimate of the damping constant $\gamma$ here, but we postpone that until the next section.

 Notice  that the very existence of a oscillatory movement would be enough to make
  the experiment critical with respect to the Abraham force. The Minkowski tensor does not predict there to be an azimuthal movement at all.

%%%%%%%%%%%%%%%%%%%%%%%%%%%%%%%%%%%%%%%%%%%%%%%%%%%%%%
%%%%%%%%%%%%%%% S E C T I O N %%%%%%%%%%%%%%%%%%%%%%%%
%%%%%%%%%%%%%%%%%%%%%%%%%%%%%%%%%%%%%%%%%%%%%%%%%%%%%%
  \section{Spherical geometry}\label{sec:sphere}

  As mentioned above, whispering gallery modes are usually associated with microspheres. Let the radius of the sphere be denoted by $a$. As above, we look for the eigenmodes,  and  we will for simplicity  focus on the TE modes only. (The meaning of the symbol  TE is here that the electric field is transverse to the radius vector $\bf r$.) We introduce quantities $\alpha$ and $\tilde r$ defined by
  \begin{equation}
  \alpha =\omega a/c, \quad \tilde{r}=r/a. \label{29}
  \end{equation}
  Thus $\alpha$ is the magnitude of the nondimensional wave vector in the exterior region (vacuum), whereas $\tilde{r}=1$ at the boundary. Making  use of the Riccati-Bessel function
  \begin{equation}
  \psi_l(x)=xj_l(x), \label{30}
  \end{equation}
  the basic TE modes in the interior can conveniently be
  written as
  \begin{subequations}
  \begin{align}
  E_r=&0, \label{31}\\
  E_\theta=&-\frac{imA_{lm}}{n\alpha \tilde{r}} \,\frac{P_l^m(\cos \theta)}{\sin \theta}\psi_l(n\alpha \tilde{r})\,F_m, \label{32}\\
  E_\phi=&\frac{A_{lm}}{n\alpha \tilde{r}} \, \frac{dP_l^m(\cos\theta)}{d\theta}\psi_l(n\alpha \tilde{r})\, F_m, \label{33}
  \end{align}
  \end{subequations}
  and
  \begin{subequations}
  \begin{align}
  H_r=&-\frac{l(l+1)}{i\omega \mu}\frac{A_{lm}}{n\alpha\tilde{r}^2}\,\frac{1}{a}P_l^m(\cos\theta)\psi_l(n\alpha \tilde{r})\,F_m, \label{34}\\
  H_\theta=&-\frac{1}{i\omega \mu} \frac{A_{lm}}{
  \tilde{r}}\,\frac{1}{a}\frac{dP_l^m(\cos
  \theta)}{d\theta} \,\psi_l'(n\alpha \tilde{r})F_m,
  \label{35}\\
  H_\phi=&-\frac{m}{\omega \mu \sin\theta}\,\frac{A_{lm}
  }{
 \tilde{r}}\,\frac{1}{a}\, P_l^m(\cos \theta)\,\psi_l'(n\alpha
  \tilde{r})F_m, \label{36}
  \end{align}
  \end{subequations}
  where $A_{lm}$ are constants, and
  \begin{equation}
  F_m=e^{im\phi-i\omega t}. \label{36a}
  \end{equation}
  The mode expansions above
  essentially follow Stratton \cite{stratton41}.

  The components of Poynting's vector are, when averaged over an optical period,
  \begin{subequations}
  \begin{align}
  S_r=&\frac{1}{2}\Re[E_\theta H_\phi^* -E_\phi H_\theta^*], \label{37}\\
  S_\theta=&\frac{1}{2}\Re [E_\phi H_r^*], \label{38}\\
  S_\phi=&-\frac{1}{2}\Re [E_\theta H_r^*]. \label{39}
  \end{align}
  \end{subequations}
  Assume that the sphere is fed by an incident flux from the
  outside such that only the component $S_\phi$ of $\bf S$ in
  the interior is different from zero. With an intensity modulated
  energy flux such as above, $S_\phi=S_0\cos \omega_0t$, we thus
  get for the azimuthally directed Abraham force density in the interior
  \begin{equation}
  f_\phi^\rmA=-\frac{n^2-1}{c^2}\omega_0S_0\sin \omega_0 t.
  \label{40}
  \end{equation}
  From the above expressions,
  \begin{equation}
  S_0=\frac{m}{2(n\alpha)^2\,\tilde{r}^3}\, \frac{l(l+1)}{\omega \mu}\,\frac{|A_{lm}|^2}{a}\frac{[P_l^m]^2}{\sin\theta}\,\psi_l^2. \label{41}
  \end{equation}
  The  Abraham torque, directed along the $z$ axis,  then becomes
  \begin{equation}
  N_z^\rmA=\int ({\bf r\times f}^\rmA)_z dV=\int rf_\phi^\rmA\sin \theta dV, \label{42}
  \end{equation}
  where the integration is over the sphere, with  $dV=r^2\sin\theta dr d\theta d\phi$. Making use of Eqs.~(\ref{40}) and (\ref{41}) we  obtain
\begin{equation}
  N_z^\rmA=-\frac{n^2-1}{c^2}\frac{\pi ma^3}{(n\alpha)^2}\frac{l(l+1)}{\omega\mu}|A_{lm}|^2\omega_0\, K_\mathrm{I}K_\mathrm{II}\,\sin \omega_0t, \label{45}
  \end{equation}
where $K_\mathrm{I}$ and $K_\mathrm{II}$ are the integrals
\begin{subequations}
  \begin{align}
    K_\mathrm{I}=& \int_0^1\psi_l^2(n\alpha \tilde{r})d\tilde{r}\notag \\
    =&\frac{1}{2}\left[ \psi_l^2(n\alpha)-\psi_{l-1}(n\alpha)\psi_{l+1}(n\alpha) \right], \label{46}\\
    K_\mathrm{II}=& \int_0^\pi[P_l^m(\cos\theta)]^2\sin \theta d\theta\notag \\
    =&\frac{2}{2l+1}\frac{(l+m)!}{(l-m)!}. \label{47}
  \end{align}
\end{subequations}
We want to relate this to the total power $P$ flowing in the azimuthal direction in the sphere. We calculate $P$ by integrating $S_\phi$ over the area of a semicircle with radius $a$,
\begin{align}
  P=&\int_0^\pi d\theta \int_0^a rdr S_\phi\notag \\
  =&\frac{ma}{2(n\alpha)^2}\frac{l(l+1)}{\omega\mu}|A_{lm}|^2 K_\mathrm{III}K_\mathrm{IV}\cos \omega_0t, \label{48}
\end{align}
where
\begin{subequations}
  \begin{align}
    K_\mathrm{III}=&\int_0^1\frac{d\tilde r}{\tilde{r}^2}\psi_l^2(n\alpha \tilde r), \label{49}\\
    K_\mathrm{IV}=&\int_0^\pi \frac{[P_l^m(\cos \theta)]^2}{\sin \theta}d\theta. \label{50}
  \end{align}
\end{subequations}
As before, it is assumed that the supplied power is intensity modulated, $P=P_0\cos \omega_0t$.

The two last integrals can be processed further, at least approximatively. First, we can rewrite $K_\mathrm{III}$ as
\begin{equation}
  K_\mathrm{III}= \frac{1}{2}\pi n\alpha \int_0^{n\alpha} \frac{dx}{x}J_\nu^2(x), \label{51}
\end{equation}
with $\nu=l+1/2$. For actual physical values, $n\alpha \gg 1$. We can thus replace the upper limit with infinity, and make use of formula 6.574.2 in Ref.\ \cite{BookGradshteyn80} to get
\begin{equation}
  K_\mathrm{III} \approx \frac{\pi n\alpha}{2(2l+1)}. \label{52}
\end{equation}
Finally, the integral $K_\mathrm{IV}$ is simply (cf. formula 8.14.14 in Ref.\ \cite{BookAbramowitz64})
\begin{equation}
  K_\mathrm{IV}=\frac{(l+m)!}{m(l-m)!}.
\end{equation}
We are now able to relate the torque $N_z^\rmA$ to the power $P$. The result becomes quite simple:
\begin{align}
  N_z^\rmA=&-\frac{n^2-1}{c^2}\,\frac{4ma^2\omega_0}{n\alpha}P_0\sin \omega_0t \notag \\
  & \times[\psi_l^2(n\alpha)
  -\psi_{l-1}(n\alpha)\psi_{l+1}(n\alpha)]. \label{53}
\end{align}
The radius of the sphere is seen to appear in the prefactor $a^2$,
as well as in the nondimensional parameter $\alpha=\omega a/c$.
The parameter $l$ occurs only as an order parameter in the
function $\psi_l$. We see that the torque is proportional to $m$.
This is as we would expect, as the whispering gallery modes are
associated with $m=l$, i.e. the maximum value of $m$. It should
correspond to a maximum angular momentum and accordingly a maximum
torque.

To proceed quantitatively, the value of $\alpha$ has to be
determined. For the TE modes it is determined by the dispersion
relation \cite{stratton41}
\begin{equation}
\frac{n\mu_0}{\mu}\frac{\psi_l'(n\alpha)}{\psi_l(n\alpha)}=\frac{{\xi_l^{(1)}}'(\alpha)}{\xi_l^{(1)}(\alpha)},
\label{54}
\end{equation}
where $\xi_l^{(1)}(x)=xh_l^{(1)}(x)$ is another member of the Riccati-Bessel functions.
The equation (\ref{54}) is complex and does not in general have real solutions, but approximate solutions with only a small imaginary inequality are found close to $\alpha \approx l$ for $l\gg 1$.

As at the end of the previous section, we focus now attention on the magnitude of the angular deflection $\phi$, as this is most likely the quantity of main experimental interest. Without changing the notation we write the Abraham torque in the form $N_z^\rmA=K\omega_0\sin \omega_0t$ as before, where now
\begin{align}
  K=&-\frac{n^2-1}{c^2}\,\frac{4ma^2}{n\alpha} \notag \\
 &\times [\psi_l^2(n\alpha)
  -\psi_{l-1}(n\alpha)\psi_{l+1}(n\alpha)]P_0. \label{55}
\end{align}
The equation of motion for $\phi$ takes the same form (\ref{25}) as before, where now the moment of inertia is
\begin{equation}
I = \frac{2}{5}Ma^2=\frac{8\pi}{15}\rho a^5, \label{56}
\end{equation}
$M$ being the mass of the sphere. For definiteness let us take $a=100~\mu$m. Then, with $\rho \sim 10^3~\rm kg/m^3$ we get $M \approx 4~\mu$g and so, with $\kappa \sim 10^{-9}~\rm N~m/rad$ as before,
\begin{equation}
\Omega \sim 10^8\sqrt{\kappa} \sim 10^3~\rm rad~s^{-1}. \label{57}
\end{equation}
With these numerical choices the value of $\Omega$ becomes of the same order as in the cylinder case. The magnitude $\phi_\text{max}$ of the maximum deflection at resonance $\omega_0=\Omega$ is now
\begin{align}
\phi_\text{max}=&\frac{10m}{Mn\alpha}\frac{n^2-1}{c^2} \notag \\
&\times [\psi_l^2(n\alpha)-\psi_{l-1}(n\alpha)\psi_{l+1}(n\alpha)]\frac{P_0}{\gamma}. \label{58}
\end{align}
As we have assumed $l\gg 1$ and $n\alpha \gg 1$ but otherwise left the ratio of these quantities unspecified, the $\psi_l$ functions ought  to be calculated numerically.

Let us finally make an estimate of the magnitude of the damping coefficient $\gamma$, assuming for definiteness that the damping is due to the viscosity of air only. We then need to know the viscous torque on a sphere executing rotary oscillations about its symmetry axis. The solution of this problem  is shown  in Ref.~\cite{landau87}. An important parameter in this context is the penetration depth $\delta ={\sqrt{2\nu/\Omega}}$, where $\nu$ is the kinematic viscosity of the surrounding medium. For air, $\nu=1.5\times 10^{-5}~\rm m^2/s$.  Thus with $\Omega \sim 10^3~\rm rad~ s^{-1}$ we get $\delta \sim 170~\upmu$m, which is of the same order as $a$.  Strictly speaking we should therefore have to use  the complete expression for the viscous torque, which is somewhat complicated. For our order-of-magnitude considerations it is however sufficient to use the simple expression
\begin{equation}
(N_z)_\text{viscous} \approx 8\pi \eta a^3\Omega, \label{59}
\end{equation}
(corresponding mathematically to the $a/\delta \ll 1$ limit), where $\eta =1.8 \times 10^{-5}\rm Pa~s$ is the dynamic viscosity for air. Identifying
 $(N_z)_\text{viscous}$ with $I\gamma \,\Omega$ in accordance with  Eq.~(\ref{25}), we  get for the damping coefficient
\begin{equation}
\gamma =\frac{8\pi \eta}{I}a^3 \sim 30~\rm s^{-1}, \label{60}
\end{equation}
and the expression (\ref{58}) for the maximum deflection can finally be written as
\begin{align}
&\phi_\text{max}=\frac{m}{2\pi n\alpha \eta a}\frac{n^2-1}{c^2} \notag \\
&\times
[\psi_l^2(n\alpha)-\psi_{l-1}(n\alpha)\psi_{l+1}(n\alpha)]{P_0}.
\label{61}
\end{align}
As expected, the deflection is very small. Whereas numerical evaluation of the $\psi_l$ functions in general is called for, as mentioned, we may note that in cases where $l \ll n\alpha$ the approximation $\psi_l(n\alpha) \approx \sin(n\alpha -\frac{l\pi}{2})$ is useful. One can moreover obtain a simple
estimate of the magnitude in the cylinder case by inserting $\gamma$ from
Eq.~(\ref{60}) into Eq.~(\ref{27}), whereby one finds $\phi_\text{max}
\sim 10^{-8}~$rad. Careful adjustments of input parameters are obviously
needed if the effect is to be verified experimentally.

\section{On the magnitude of torques in existing experiments}

We close this investigation by making some estimates of radiation torques on spheres, as well as  on  ring resonators (a closely related geometry),  for already existing experiments. As first example we take the setup reported in
 Ref.\ \cite{spillane02}, where an infrared laser of wavelength $\lambda=1500~$nm was used. Two different sphere radii were investigated, $a=40~\upmu$m and $a=70~\upmu$m, corresponding to values of $\alpha\approx l=m$ equal to  $162$ and $283$, respectively. Although the feeding laser had a power in the order of tens of microwatts to milliwatts, the extremely high $Q$ factor of the silica sphere meant the buildup of circulating modes in the sphere grew enormous. Circulating powers in excess of $100$W are routinely reported in such systems (e.g.\ \cite{rokshari05}) (although this quantity was not explicitly given in the reference \cite{spillane02}). The refractive index of materials used for ultra-high-$Q$ spherical resonators, such as fused silica \cite{spillane02,gorodetsky98} and quartz \cite{vernooy98}, are about $n=1.5$. With these values as input for $P_0$ we obtain the torques [$N_z^\rmA=N_0 \sin \omega_0t$]
\begin{equation}\label{numest}
  N_0 \approx \left\{ \begin{array}{c}4\times 10^{-24}\text{Nms} \cdot \omega_0\\ 1\times 10^{-23}\text{Nms} \cdot \omega_0\end{array}, \right. \text{ for }  a=\left\{ \begin{array}{c}40\upmu\text{m} \\ 70\upmu\text{m}\end{array}\right. .
\end{equation}
Note in general  that for a sphere, $N_z^\rmA \propto a$ according to Eq.~(\ref{53}), whereas  $\phi_\text{max} \propto a^{-2}$ according to Eq.~(\ref{61}) when the viscous damping is accounted for.

The geometry of Ref.\ \cite{rokshari05}, which reports circulating powers in excess of $100$ W, employs the  toroidal ring resonator. This geometry has the benefit of having smaller mass and therefore smaller moment of inertia than a sphere of the same radius, allowing for larger angular deflections. For a thin ring the moment of inertia is
\be
  I_\text{toroid} \approx 2\pi\rho A a^3
\ee
where $A$ is the area of cross-section. The torque on such a toroid would be roughly similar to that on a sphere, so it is reasonable to assume the angular deflection to be larger and scale as $a^{-1}$. This could allow larger radii which could be beneficial for detection. Cf. also the review article \cite{vahala02}.

We wish finally to re-emphasize the possibility of using quite high frequencies $\omega_0$ in order to produce measurable values for the Abraham torque.  We  assumed above the strong field inside the microcavity to react instantaneously to the sinusoidal variations of the input signal, an approximation which is good provided the build-up and ringdown time ($\tau$) of the resonator is small compared to $2\pi/\omega_0$. For the $45\upmu$m radius toroidal resonator in Ref.\ \cite{armani03}, for example, a ringdown time of about $43$ ns was measured. For  cavities of even higher $Q$-factor, ringdown times are somewhat longer, yet this implies that we may choose tuning frequencies $\omega_0$ as high as $10^6$ without invalidating the theory. Due to the proportionality of the torque with $\omega_0$, going close to the megahertz regime could increase the torque to perhaps $10^{-17}~$Nm for a sphere with radius of some tens of microns.

\bigskip

{\bf Acknowledgment}

\bigskip
I. B. thanks Giovanni Carugno for correspondence concerning measurement accuracies.

\end{document}